\documentclass[onecollarge]{svjour2}       % onecolumn "king-size"
\smartqed  % flush right qed marks, e.g. at end of proof
\usepackage{graphicx}
\usepackage{amssymb}
\begin{document}

\title{The WITCH Experiment: towards weak interactions studies. Status and prospects%\thanks{Grants or other notes
}


\author{V.Yu.~Kozlov\footnote{\emph{E-mail address:} Valentin.Kozlov@cern.ch}  \and
        M.~Beck \and
        S.~Coeck \and
        M.~Herbane \and
        I.S.~Kraev \and
        N.~Severijns \and
        F.~Wauters \and
        P.~Delahaye \and
        A.~Herlert \and
        F.~Wenander \and
        D.~Z\'{a}kouck\'{y} \and
        (ISOLDE, NIPNET and TRAPSPEC collaborations) %etc.
}


\institute{V.Yu.~Kozlov, M.~Beck, S.~Coeck, M.~Herbane,
I.S.~Kraev, N.~Severijns and F.~Wauters \at
               K.U.Leuven, Instituut voor Kern- en Stralingsfysica, Celestijnenlaan~200D, \mbox{B-3001~Leuven}, Belgium \\
%              Tel.: +123-45-678910\\
%              Fax: +123-45-678910\\
              \emph{Present address of M. Beck:} Universit\"{a}t M\"{u}nster, Institut f\"{u}r Kernphysik, Wilhelm-Klemm-Str. 9, D-48149
             M\"{u}nster, Germany %  if needed
%              \email{Valentin.Kozlov@cern.ch}           %  \\
           \and
%           M. Beck \at
%              Universit\"{a}t M\"{u}nster, Institut f\"{u}r Kernphysik, Wilhelm-Klemm-Str. 9, D-48149
%             M\"{u}nster, Germany
           \and
           P.~Delahaye, A.~Herlert and F.~Wenander \at
           CERN, CH-1211~Gen\`eve 23, Switzerland
           \and
           D.~Z\'{a}kouck\'{y} \at
           Nuclear Physics Institute, ASCR, 250 68 \v{R}e\v{z}, Czech Republic
}

\date{Received: date / Accepted: date}

\maketitle

\begin{abstract}
Primary goal of the WITCH experiment is to test the Standard Model
for a possible admixture of a scalar or tensor type interaction in
$\beta$-decay. This information will be inferred from the shape of
the recoil energy spectrum. The experimental set-up was completed
and is under intensive commissioning at ISOLDE (CERN). It combines
a Penning trap to store the ions and a retardation spectrometer to
probe the recoil ion energy. A brief overview of the WITCH set-up
and the results of commissioning tests performed until now are
presented. Finally, perspectives of the physics program are
reviewed. \keywords{Weak interactions \and Penning trap \and
Retardation spectrometer} \PACS{23.40.Bw \and 24.80.+y \and
29.25.Rm \and 29.30.Ep}
% PACS:
% 12.60.Cn    Extensions of electroweak gauge sector
% 23.40.Bw    Weak-interaction and lepton (including neutrino) aspects (see also 14.60.Pq Neutrino mass and mixing)
% 24.80.+y    Nuclear tests of fundamental interactions and symmetries
% 29.25.Rm    Sources of radioactive nuclei
% 29.30.Ep    Charged-particle spectroscopy
\end{abstract}

\section{Introduction}
\label{intro} The most general interaction Hamiltonian for nuclear
$\beta$-decay which includes all possible interaction types
consistent with Lorentz-invariance \cite{lee56,jackson57a}
contains 5 different terms, so-called Scalar (S), Vector (V),
Tensor (T), Axial-Vector (A) and Pseudoscalar (P) contributions.
The Standard Model (SM) of the weak interaction excepts only V and
A interactions which leads to the well-known $V-A$ structure of
the weak interaction. However, the presence of scalar and tensor
types of weak interaction is today ruled out only to the level of
about 8\% of the V- and A-interactions~\cite{severijns05a}.

A possible admixture of a scalar or tensor type weak interaction
in $\beta$-decay can be studied by determining the $\beta-\nu$
angular correlation. This correlation for unpolarized nuclei can
be characterized by the $\beta-\nu$ angular correlation
coefficient \emph{a}~\cite{jackson57b} which is
%%%%%%%%%%%%%%%%%%%%%%%%%%%%%%%%%%%%%%%%
%written as~\cite{jackson57b}:
%
%\begin{equation}
%\omega(\theta_{\beta\nu})\backsimeq1+a\cdot\frac{v_{\beta}}{c}\cos\,\theta_{\beta\nu}\left[1-\frac{\Gamma
%m}{E}b\right] \label{eq:beta-nu-angl}
%\end{equation}
%
%\noindent where $\theta_{\beta\nu}$ is the angle between the
%$\beta$ particle and the neutrino, \emph{E}, $v_{\beta}/c$ and
%\emph{m} are the total energy, the velocity relative to the speed
%of light and the rest mass of the $\beta$ particle,
%$\Gamma=\sqrt{1-(\alpha Z)^{2}}$ with $\alpha$ the fine-structure
%constant and $Z$ the nuclear charge of the daughter nucleus,
%\emph{b} is the Fierz interference term \emph{}which has
%experimentally been shown to be small (e.g.
%$\left|b_{F}\right|<0.0044$ at 90\%~C.L.~\cite{hardy05}) and can
%as a first approximation be assumed to be zero, and \emph{a} is
%the $\beta-\nu$ angular correlation coefficient which is for
%%%%%%%%%%%%%%%%%%%%%%%%%%%%%%%%%%%%%%%%%%%%%%%%%%%%%%%%%%%%%%%%%%%
for instance for pure Fermi transitions given by:

\begin{figure*}[!t]
\centering
  \includegraphics[width=0.55\textwidth]{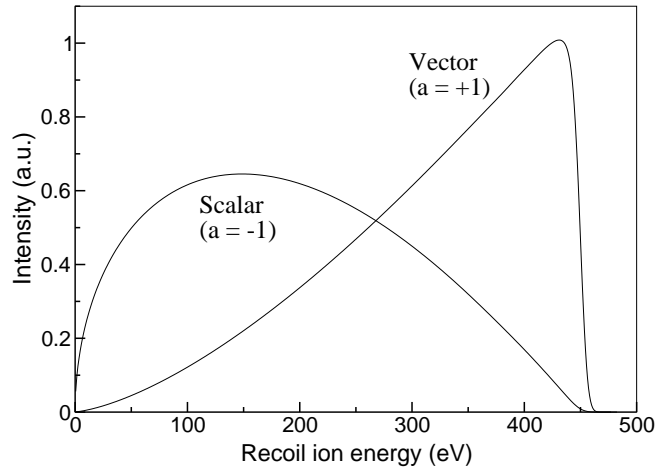}
\caption[Differential recoil energy spectrum]{Differential recoil
energy spectrum for $a=1$ (pure V interaction) and $a=-1$ (pure S
interaction) .}
\label{fig:1-recoils-v-s}       % Give a unique label
\end{figure*}
\begin{eqnarray}
a_{F} & = &
\frac{\left|C_{V}\right|^{2}+\left|C_{V}^{\prime}\right|^{2}-\left|C_{S}\right|^{2}-\left|C_{S}^{\prime}\right|^{2}}{\left|C_{V}\right|^{2}+\left|C_{V}^{\prime}\right|^{2}+\left|C_{S}\right|^{2}+\left|C_{S}^{\prime}\right|^{2}}\,
\label{eq:aF}
\end{eqnarray}

Since the Standard Model assumes only V- and A- interactions
$a^{SM}_{F}=1$ ($a^{SM}_{GT}=-1/3$ for pure Gamow-Teller). Any
admixture of S- to V- (T to A) interaction in a pure Fermi
(Gamow-Teller) decay would result in $a<1$ ($a>-1/3$). It can be
shown that the two leptons in $\beta$-decay will be emitted
preferably into the same direction for a \emph{}V~(T) interaction
and into opposite directions for an S~(A) interaction leading to a
relatively large energy of the recoil ion for a V~(T) interaction
and a relatively small recoil energy for an S~(A) interaction
(Fig.~\ref{fig:1-recoils-v-s}). The WITCH experiment as a primary
goal aims to measure the shape of the recoil energy spectrum in
nuclear $\beta$-decay with high precision in order to deduce the
parameter \emph{a} which will give information on a possible
scalar or tensor type interaction. Similar experiments were
recently performed at TRIUMF \cite{gorelov05} and at Berkley
\cite{scielzo04}, while another experiment is ongoing at GANIL
\cite{rodriguez05}.

\section{Experiment}
\label{sec:1}
\begin{figure*}
\centering
  \includegraphics[width=0.5\textwidth]{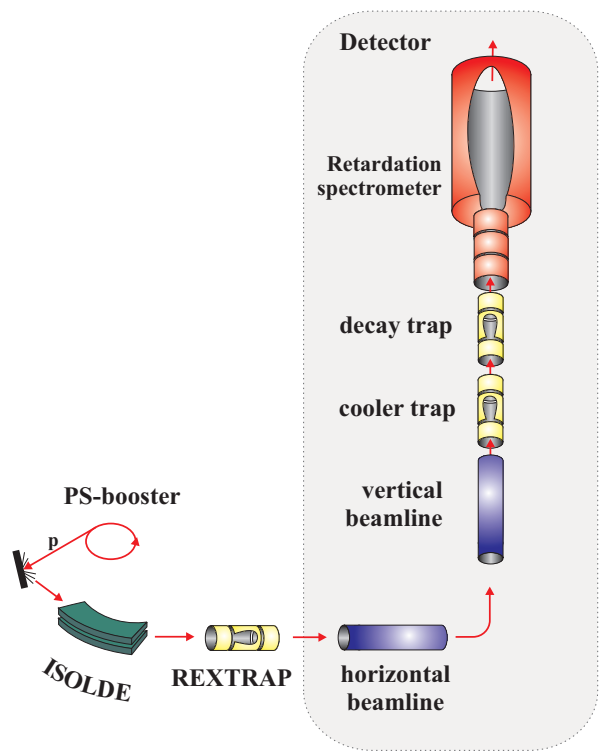}
\caption[WITCH set-up]{General schematic view of the WITCH set-up
(not to scale).}
\label{fig:2-setup}       % Give a unique label
\end{figure*}

%\subsection{Experimental set-up}\label{subsec:1.1}
In order to fulfill the goal of the experiment the WITCH set-up
was built and installed in ISOLDE (CERN). The main feature of this
set-up is a combination of a double Penning trap structure to
store the $\beta$-decaying radioactive ions and a retardation
spectrometer to probe the energy of the daughter recoil ion
(Fig.~\ref{fig:2-setup}).

The radioactive ions are produced by the ISOLDE facility
\cite{kugler00} and are then first trapped, cooled and bunched by
REXTRAP~\cite{ames05}. The ion bunches ejected from REXTRAP have
60~keV (optionally 30~keV) energy and are guided into the WITCH
set-up. They are decelerated from 60~keV to about 80~eV in the
vertical beamline by means of the pulsed drift
cavity~\cite{delaure06} and thus can be trapped in the first
Penning trap, the cooler trap. This trap serves to cool the ions
further down in energy by buffer gas collisions and to prepare the
ion cloud to be injected into the second Penning trap, the decay
trap. Both traps are placed in a 9~T magnetic field. After
$\beta$-decay ions leave the decay trap and those emitted in the
direction of the spectrometer (upper half hemisphere) are probed
for their energy by an electrostatic retardation potential. The
magnetic field in the region of the retardation analysis plane is
0.1~T. The working principle of the retardation spectrometer is
based on the principle of adiabatic invariance of the magnetic
flux and similar to the $\beta$-spectrometers used for the
neutrino rest-mass measurements. The ions that pass the
retardation plane are re-accelerated to $\sim$10~keV and focused
onto the micro-channel plate (MCP) detector which is equipped with
 delay line anodes for position sensitivity. Changing the
retardation potential over the necessary range allows to measure a
recoil ion spectrum with high precision.

Simulations performed show that for a reasonable measurement time
one can reach a precision for the $\beta-\nu$ angular correlation
parameter $\Delta a = 0.005$ which corresponds to
$\left|C_{S}\right|\lesssim9\%$~\footnote{More detailed
information on scalar constraints can be found
in~\cite{adelberger99,severijns05a}} at C.L.=95\%. Further
improvements in order to push the limit down to $\Delta a =
0.002$, or $\left|C_{S}\right|\lesssim6\%$ at C.L.=95\%, should be
possible.

More details about the WITCH set-up can be found in
\cite{beck03a,kozlov06}.

\section{Present status}
\label{sec:2} Tests performed to check the beam transport, to
confirm the functionality of pulsing down the ions and to verify
the basic operation of the Penning traps are described in
\cite{kozlov06,delaure06}. Here we present only the most recent
results.
\subsection{Traps}\label{subsec:traps}
\begin{figure*}
\centering
  \includegraphics[width=0.65\textwidth]{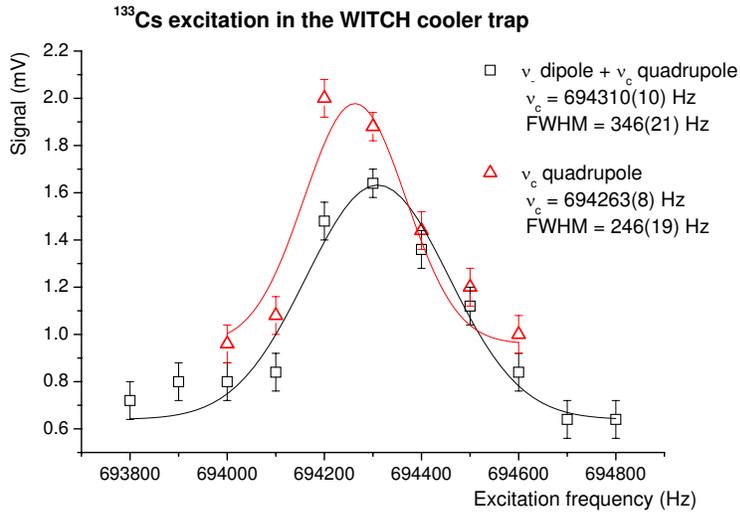}
\caption[Quadrupole excitation]{Cooling resonances for $^{133}$Cs
as obtained from the MCP ion signal as a function of the applied
frequency. Ions are ejected from the cooler trap. The B-field was
6~T.}
\label{fig:3-omega-c}       % Give a unique label
\end{figure*}
The main purpose of the cooler trap is to prepare the ion cloud
for the measurement. This involves buffer gas cooling, cleaning
from possible contaminates, better centering and cooling of the
magnetron motion to prevent the cloud from expanding. The last
three tasks can be achieved by \emph{sideband cooling}
\cite{konig95}. This technique was tested by measuring the signal
on the MCP detector installed in the middle of the retardation
spectrometer as a function of the applied frequency near the true
cyclotron frequency. The isotope used is $^{133}$Cs. In a first
step all ions in the cooler trap were driven out of the center
with dipole excitation and then the quadrupole excitation was
applied (Fig.~\ref{fig:3-omega-c}, square data points). Only ions
which can be re-centered can reach the detector because they have
to pass the 3~mm diameter pumping diaphragm (see
Fig.~\ref{fig:6-traps-diaph}). As can be seen from
Fig.~\ref{fig:3-omega-c} a nice resonance curve is obtained with
$\Delta\nu$(FWHM)=346~Hz corresponding to a mass resolving power
$R=\nu/\Delta\nu$(FWHM)$\approx$2000. This is enough to completely
remove isotopic contaminations which can be present in the
incoming beam. A pure quadrupole excitation without prior dipole
excitation was applied as well. As one can see from
Fig.~\ref{fig:3-omega-c}~(triangle data points) it has a similar
effect as sideband cooling meaning that the initial injection into
the cooler trap was not ideal, e.g. ions were injected off center
and/or they had a too large magnetron component. However, using
the quadrupole excitation still allowed to correct for this.

It has to be noted here that the magnetic field homogeneity allows
to improve the mass resolving power up to $R=10^4\div10^5$.

\subsection{First recoil ions}\label{subsec:recoils}
\begin{figure*}
\begin{minipage}[c]{1.0\columnwidth}%
\centering\includegraphics[%
  width=0.49\textwidth,
  keepaspectratio]{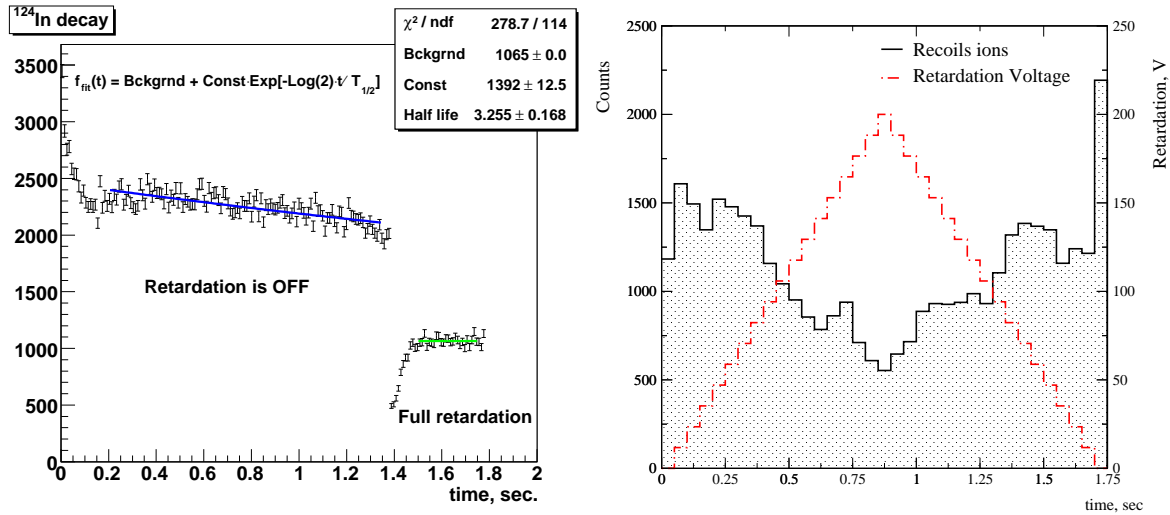}~
  \includegraphics[width=0.49\textwidth,
  keepaspectratio]{figs/fig4_124In_retard.eps}\end{minipage}%
\caption[Retardation spectrum]{Retardation spectra for
$^{124g,m}$In $\beta$-decay. \emph{Left}: simple On-Off
measurement; \emph{Right}: 35 retardation steps are used in total,
first rising from 0 to 200~V then going back to 0~V (retardation
steps are indicated).}
\label{fig:4-retard}       % Give a unique label
\end{figure*}
In June 2006 the WITCH experiment observed first recoil ions
following the $\beta$-decay of trapped $^{124}$In
(Fig.~\ref{fig:4-retard}). The incoming beam was a mixture of
$^{124g}$In ($t_{1/2}$=3.11~s) and $^{124m}$In ($t_{1/2}$=3.7~s).
As a first step a simple on-off measurement was performed: for
1.4~s no retardation was applied thus measuring the total signal
and then for 0.4~s 400~V retardation was used stopping all recoil
ions and thus providing the background signal
(Fig.~\ref{fig:4-retard}, left). As expected the count rate drops
down when the retardation is On. The data were fitted and a
half-life of 3.3(3)~s was found which is in a good agreement with
the incoming mixed beam of $^{124g,m}$In. The latter confirms that
the ions trapped in the decay trap are $^{124g,m}$In. As a next
step different retardation schemes were used. However, strong
discharge in the re-acceleration part of the set-up (see
Sect.~\ref{sec:1}) was observed. Thus the acceleration voltages
were significantly reduced which, however, caused a focusing
effect, i.e. recoil ions were focused differently depending on
their energy and could miss the detector. Nevertheless, a correct
general behavior of the spectrometer was observed
(Fig.~\ref{fig:4-retard}, right): increasing the retardation leads
to lower count rate on the detector while decrease of the
retardation corresponds to higher count rate.

\subsection{Position sensitivity}\label{subsec:mcp}
%\begin{figure*}
%\begin{minipage}[c]{1.0\columnwidth}%
%\centering\includegraphics[%
%  width=0.49\textwidth,
%  keepaspectratio]{figs/fig5_tx1_tx2.eps}~
%  \includegraphics[width=0.49\textwidth,
%  keepaspectratio]{figs/fig5_ty1_ty2.eps}\end{minipage}%
%\caption[MCP XY delay sum]{Sum of the arrival times on both ends
%of the X (left) and Y (right) lines.}
%\label{fig:5-mcp-xy}       % Give a unique label
%\end{figure*}
Recently a position sensitive MCP of 4~cm diameter was installed
as the main detector in the system. This MCP allows to detect a
larger beam spot and to study a number of systematic effects like
a possible dependence of the beam size on the ion energy, the
focusing problem which already played a role during the beam time
(see Sect.~\ref{subsec:recoils}) and the effect of the
$\beta$-background. This MCP is of the same type as described in
\cite{lienard05}. Position sensitivity is realized by delay line
anodes, i.e. the one-dimensional position of the ion hit is
deduced from the difference of the propagation times to both ends
of the corresponding wire. The position sensitivity read-out was
tested by installing a special mask in front of the detector.
Reading back the timing signals and reconstructing the hit
positions allowed to reproduce the shape of the mask, thus
confirming the functionality of the system. Presently achieved
position resolution is of the order of 1~mm. This MCP is located
in the magnetic field of about 30~G, but no dependency of the MCP
resolution on B-field was studied yet.

\section{Outlook} \label{sec:3}
\begin{figure*}
\centering
  \includegraphics[width=0.75\textwidth]{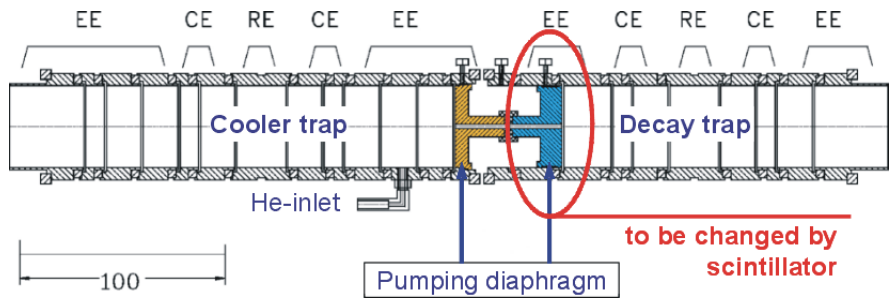}
\caption[WITCH traps]{WITCH double Penning trap structure. Part of
the pumping diaphragm to be changed is indicated.}
\label{fig:6-traps-diaph}       % Give a unique label
\end{figure*}
As for technical improvements we currently investigate the
possibility to change part of the pumping diaphragm to a
$\beta$-detector (optionally to a scintillator
detector)(Fig.~\ref{fig:6-traps-diaph}). This will allow to have
an additional normalization between different trap loads, to have
a start signal for TOF measurements and to evaluate the
$\beta$-background on the main MCP detector on-line.

The WITCH physics program is not limited to the search for exotic
scalar or tensor currents. If one assumes that the Standard Model
is correct measuring the $\beta-\nu$ correlation parameter
\emph{a} yields information about the Fermi to Gamow-Teller mixing
ratio. Due to the properties of the retardation spectrometer
different charge states will appear at different positions in the
spectrum thus allowing to study the charge state distribution
after nuclear $\beta$-decay. Also, ions from $\beta$-decay and
from electron capture (EC) can be separated in the WITCH
spectrometer. This opens the possibility to determine the
EC/$\beta^+$-branching ratio. However, for this one has to include
the charge state distribution as a parameter and has to measure
the whole recoil spectrum. One can also study the possibility to
search for heavy neutrinos. The presence of these neutrinos can be
indicated by subtle kinks in the recoil spectrum.

\section{Conclusion} \label{sec:4}

The WITCH experiment has as the primary goal to measure exotic
scalar or tensor currents in the weak interaction. The set-up is
installed in ISOLDE (CERN). All major components of the set-up
including beamline transport, double Penning trap system and the
retardation spectrometer are checked for their functionality and
show good or acceptable efficiency (Table~\ref{tab:1-eff}). The
overall efficiency of the set-up was improved by up to a factor of
30 in comparison with 2004 (see \cite{kozlov06}). This allowed to
detect first recoil ions in the WITCH set-up. A number of
systematic effects still has to be studied and certain
improvements have to be made in order to further increase the
efficiency and to perform precise measurement of the recoil ion
spectrum.

\begin{table}[!h]
\caption{Efficiencies for the ideal WITCH set-up and currently
best achieved values (for parameters that were not yet studied the
values of an ideal set-up are taken to calculate the total
efficiency).}
\label{tab:1-eff}       % Give a unique label
\centering
\begin{tabular}{lcc}
\hline \rule{0pt}{10pt}\textbf{Description}& \textbf{ideal}&
\textbf{Best achieved}\tabularnewline & \textbf{set-up}&
\textbf{2004-2006}\tabularnewline \hline \rule{0pt}{10pt}Beamline
transfer + pulse down& 50\%& $\sim$50\%\tabularnewline Injection
into B-field& 100\%& 10\%\tabularnewline  Cooler trap efficiency&
100\%& $\sim$60\%\tabularnewline  Transfer between traps& 100\%&
$\sim$80\%\tabularnewline  Storage in the decay trap& 100\%&
100\%\tabularnewline  Fraction of ions leaving the decay trap&
40\%& not studied\tabularnewline  Shake-off for charge state n=1&
10\%& not studied\tabularnewline  Transmission through the
spectrometer& 100\%& $\sim$50\%\tabularnewline MCP efficiency&
60\%& 52.3(3)\% \cite{lienard05}\tabularnewline \hline
\rule{0pt}{10pt}\textbf{Total efficiency}& \textbf{$\sim$1\%}&
\textbf{$\sim$2.5$\cdot$10$^{-2}$\%}\tabularnewline
\end{tabular}
\end{table}

\begin{acknowledgements}
This work is supported by the European Union grants FMRX-CT97-0144
(the EUROTRAPS TMR network) and HPRI-CT-2001-50034 (the NIPNET RTD
network), by the Flemish Fund for Scientific Research FWO and by
the projects GOA 99-02 and GOA 2004/03 of the K.U.Leuven.
\end{acknowledgements}

\bibliographystyle{spmpsci}      % mathematics and physical sciences


\end{document}